\documentclass[useAMS,usenatbib]{mn2e}

\usepackage{hyperref}
\newcommand{\adsurl}[1]{\href{#1}{ADS}}

\providecommand{\url}[1]{\href{#1}{#1}}

\def\kev{~{\rm keV}}
\def\Mpc{~{\rm Mpc}}
\def\msun{M_{\odot}}
\def\beq{\begin{equation}}
\def\eeq{\end{equation}}
\def\bea{\begin{eqnarray}}
\def\eea{\end{eqnarray}}
\def\eg{e.g.,~}

\usepackage{graphicx}
\usepackage{amssym,aas_macros}

\title{Missing Thermal Energy of the Intracluster Medium}

\author[Niayesh Afshordi, Yen-Ting Lin, Daisuke Nagai, and Alastair J. R.
Sanderson] {Niayesh Afshordi,$^1$\thanks{Electronic address:
\href{mailto:nafshordi@cfa.harvard.edu}{nafshordi@cfa.harvard.edu}}
Yen-Ting Lin$,^{2,3}$
Daisuke Nagai,$^4$ and Alastair J. R. Sanderson$^5$ \\
$^1$Institute for Theory and Computation, Harvard-Smithsonian Center
for Astrophysics, MS-51, 60 Garden Street, Cambridge, MA 02138, USA
\\ $^2$Princeton University Observatory, Princeton University,
Princeton, NJ 08544,
USA \\
$^3$Departamento de Astronom\'{i}a y Astrof\'{i}sica, Pontificia Universidad Cat\'{o}lica de Chile \\
$^4$Theoretical Astrophysics, California Institute of Technology,
Mail Code 130-33, Pasadena, CA 91125, USA\\
$^5$School of Physics and Astronomy, University of Birmingham,
Edgbaston, Birmingham B15 2TT, UK}

\date{\today}
\pagerange{\pageref{firstpage}--\pageref{lastpage}} \pubyear{2006}

\def\LaTeX{L\kern-.36em\raise.3ex\hbox{a}\kern-.15em
    T\kern-.1667em\lower.7ex\hbox{E}\kern-.125emX}

\begin{document}

\maketitle

\label{firstpage}

\begin{abstract}
  The Sunyaev-Zel'dovich (SZ) effect is a direct probe of thermal energy content
  of the Universe, induced in the cosmic microwave background (CMB) sky
  through scattering of CMB photons off hot electrons in the intracluster
  medium (ICM). We report a $9\sigma$ detection of the SZ signal in the CMB
  maps of Wilkinson Microwave Anisotropy Probe (WMAP) 3yr data, through study
  of a sample of 193 massive galaxy clusters with observed X-ray temperatures
  greater than $3\kev$.  For the first time, we make a model-independent
  measurement of the pressure profile in the outskirts of the ICM, and show
  that it closely follows the profiles obtained by X-ray observations and
  numerical simulations.  We find that our measurements of the SZ effect would
  account for only half of the thermal energy of the cluster, if all the
  cluster baryons were in the hot ICM phase.  Our measurements indicate that a
  significant fraction, $35\% \pm 8 \%$, of baryonic mass are missing from the
  hot ICM, and thus must have cooled to form galaxies, intracluster stars, or
  an unknown cold phase of the ICM.  There does not seem to be enough mass in
  the form of stars or cold gas in the cluster galaxies or intracluster space,
  signaling the need for a yet-unknown baryonic component (at 3$\sigma$ level), or otherwise new
  astrophysical processes in the ICM.
\end{abstract}


\section{Introduction}
Galaxy clusters are remarkable laboratories for studying structure formation
and cosmology.  Thanks to their physical size of $\sim \Mpc$, they can be
easily resolved in different frequencies, ranging from radio, optical to
X-rays, and thus any model for cluster physics can be independently tested
against various observations. The simple model that is borne out of these
studies can then be used as a standard candle/ruler that probes the geometry
and dynamics of the Universe at cosmological redshifts
\citep[\eg][]{borgani_etal06}.

However, the simple theoretical picture of cluster formation through
gravitational collapse is complicated by other astrophysical
processes such as shock-heating and radiative cooling of gas, star
formation, energy feedback and chemical enrichment of the ICM by
supernovae and accreting super-massive black holes.  Most of our
observational understanding of the intracluster medium (ICM) to date
has come from the study of its diffuse X-ray emission, which has led
to a multitude of theoretical investigations on the effects of gas
cooling and heating of the ICM \citep[see][for a
review]{voit_etal05}. Although modern high-resolution hydrodynamic
simulations have enabled detailed theoretical modeling of cluster
formation, the details of star formation or interactions with
super-massive black holes are still far beyond the numerical
resolution limit, and our understanding of details and relative
importance of most of these processes still remain quite uncertain.
Therefore, further observational studies of clusters in X-ray and
other wavebands will continue to be the key in guiding physical
modeling of cluster properties and formation.

The thermal Sunyaev Zel'dovich (SZ) effect is a unique observational probe of
galaxy clusters detectable in radio and microwave frequencies.  It is a
spectral distortion in CMB spectrum caused by inverse-Compton scattering of
CMB photons off hot electrons in the ICM \citep{sunyaev72,birkinshaw_etal99}.
As the number of photons does not change through scattering, the SZ effect
causes a decrement (increment) in the intensity of CMB at low (high)
frequencies.  The effect has the unique property that its signal is
independent of redshift, and it directly probes the ICM pressure, or thermal
energy density.  These features make the SZ effect a particularly unique and
powerful observational tool for detecting clusters at cosmological redshifts
($z\gtrsim 0.5$) and hence a promising cosmological probe of dark energy
\citep[see][for a review]{carlstrom_etal02}.

Although various CMB experiments are now aiming at carrying out SZ cluster
surveys, all-sky CMB surveys with sufficient resolution to detect many SZ
clusters directly \citep[such as {\it Planck};][]{2006astro.ph.11567G} are yet
non-existent. The highest resolution all-sky map of the CMB sky has been
recently released by the Wilkinson Microwave Anisotropy Probe (WMAP) group
\citep{hinshaw06}. The WMAP maps alone do not show any significant signature
of the SZ effect \citep{spergel06}.  Although the WMAP SZ signals of
individual cluster have a low significance, it is possible to combine SZ
signatures of many clusters to obtain constraints on the mean ICM properties.
First efforts in this direction have been made through direct
cross-correlation of galaxy/cluster surveys in optical/IR bands with the CMB
temperature maps obtained by WMAP
\citep{bennett_etal03,fosalba_etal03,fosalba_etal04,myers_etal04,afshordi_etal04}.
These measurements have led to 2-5$\sigma$ detection of anti-correlation with
different galaxy and cluster surveys, which is consistent with the expected SZ
signal at WMAP frequencies \citep{afshordi_etal04,hernandez_etal04}.

More optimal detection of SZ effect is possible via cross-correlation of the
CMB maps with the temperature of the X-ray emitting ICM, since both the SZ and
X-ray probe the same hot ICM.  This was a primary motivation for our study of
the WMAP 1st year data.  In \citet*[][ALS05 hereafter]{ALS}, we reported an
$8\sigma$ detection of the SZ signal, through an optimized correlation of WMAP
1st year data release \citep{bennett03} with a catalog of X-ray clusters.
Using a theoretically motivated ICM profile for the analysis, ALS05 have shown
that the gas fraction in clusters is $\sim 30-40\%$ low compared to the
universal baryon fraction of the Universe.  If the clusters contain a
representative mix of dark matter and baryons of the Universe
\citep{evrard90,metzler_evrard94,navarro_etal95,lubin_etal96,eke_etal98,frenk_etal99,mohr_etal99,bialek_etal01,kravtsov_etal05,ettori_etal06,mccarthy_etal06},
this number would be far too low to account for the expected baryon budgets in
clusters, since the observed stellar and cold gas fraction is no larger than
$\approx 10-15$\% \citep{gonzalez_etal04,lin04}.  These results are in
agreement with other SZ \citep{laroque_etal06} and X-ray
\citep{ettori_etal03,vikhlinin_etal06} studies.  These results therefore
indicate that the baryons are missing from the expected baryon budget of
clusters.

Much larger discrepancies with X-ray observations have been reported
in \citet{2006ApJ...648..176L}, who claim that the WMAP SZ signal
for ROSAT X-ray clusters is a factor of $\sim 4$ smaller than the
expectation from X-ray brightness profiles. However, in order to
compare with WMAP maps, \citet{2006ApJ...648..176L} extrapolate the
$\beta$-model fits far beyond the region that is fit by X-ray data,
suggesting that the reported discrepancy may simply be an artifact
of this extrapolation, which is known to fail at large radii
\citep[\eg][]{1999ApJ...525...47V,2004MNRAS.352.1413S,vikhlinin_etal06}.

Given the importance of the problem, we repeat the measurements of
ALS05 using the 3 year WMAP data in this paper.  The improvements
from the previous analyses are two-folds. First, the WMAP 3yr data
release \citep{hinshaw06} should provide a higher significance
detection of the SZ signal, while still capturing large angle
information that is missing from higher resolution interferometric
experiments. Second, we devise a model-independent method to measure
the mean ICM pressure profile from WMAP temperature maps for a
catalog of X-ray clusters.  The latter will help minimize the
dependence on theoretical assumptions and associated systematic
uncertainties, which often dominate systematic uncertainties in SZ
anlayses carried out using a cluster model \citep[see e.g.,
ALS05;][]{laroque_etal06}.  We then use this measurement to put
constraints on the thermal energy and baryonic budget of the ICM.

The paper is organized as follows. In the next section we describe our
methodology, including compilation of our X-ray cluster catalog, our method to
extract the SZ signal from the CMB maps, and the hydrodynamic simulations that
we compare to the data. Our main observational results are described in
Sec.~\ref{results}. In Sec.~\ref{discuss}, we discuss the possible
shortcomings, as well as the implications of our analysis, the most puzzling
of which is a missing baryonic component of the ICM.  Finally,
Sec.~\ref{sec:future} concludes the paper.



\section{Methodology}

\subsection{X-ray cluster catalog}\label{x-ray}
\begin{figure}
\includegraphics[width=\linewidth]{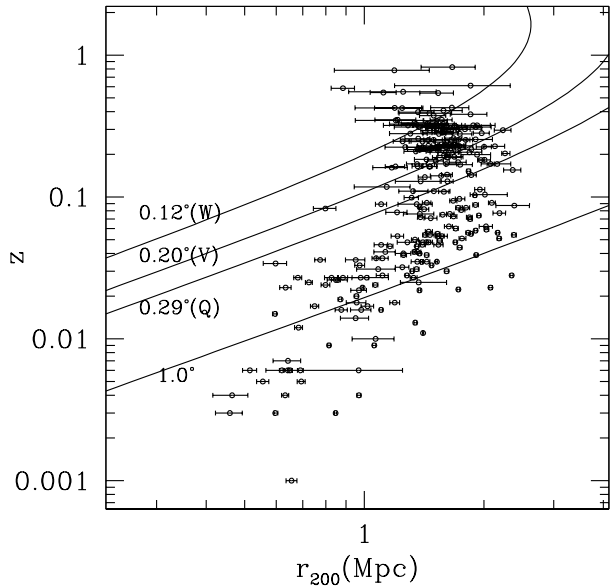}
\caption{\label{zrvir} The distribution of cluster redshifts and
virial radii (estimated from X-ray temperature; see Sec.
\ref{extract}). The three upper lines show the resolution of WMAP
bands \citep[associated with the radius of the disk with the same
effective area as the detector beams; see][]{2006astro.ph..3452J},
while the lower line shows the physical radius of the 1 degree
circle
 at the cluster redshift. }
\end{figure}

Our primary objective is to study the SZ signal in a large sample of
galaxy clusters by combining signals from known X-ray clusters.
Therefore, we first assemble X-ray cluster catalog by searching the
X-ray Galaxy Clusters Database
(BAX)\footnote{\href{http://bax.ast.obs-mip.fr/}{http://bax.ast.obs-mip.fr/}}.
The database provides a comprehensive compilation of cluster
properties from the literature, including the position on the sky,
redshift, and X-ray measurements such as luminosity and/or
temperature. The first pass in our cluster selection requires that
clusters have measured X-ray temperature. We further remove clusters
with multiple entries in the database, and those that lie too close
to each other in projection.  Although the resulting sample is not
based on any statistically complete selection, we attempt to make
the temperature measurements as consistent as possible, by adopting
the ones that do not include the central regions of the clusters, or
those obtained through a two-temperature component analysis, thus
minimizing effect of any ``cool core'' (CC) in the cluster mass
determination. Preference is given to a few studies that provide
such temperature measurement for large samples of clusters
\citep{finoguenov01,fukazawa04,ikebe02}. While in the former two
studies, the central regions have been excised, in the latter, the
temperature is given by the hotter component of their
two-temperature model.  This leaves us with a sample of 260 clusters
with measured X-ray temperatures (Fig.~\ref{zrvir}).

For a fraction of our final cluster sample, estimates of the ICM gas cooling
time in the cluster central region are available in the literature
\citep{white97,peres98,bauer05}, thus allowing an investigation of the effect
of the cool core on the cluster baryon budget. Cool cores are found in $\sim
50\%$ of nearby clusters, and although they show similar scaling properties
across a wide mass range, they behave very differently from the inner regions
of non-CC clusters \citep{sanderson06}. We refer to CC clusters as those with
their central cooling times more than $3\sigma$ below half of the Hubble time
(7 Gyr). In particular, for a subset of our sample with $T_X > 3\kev$ (and
after point source contamination cut, explained in Sec.  \ref{extract}), which
comprises 193 of our clusters, 80 clusters have measured central cooling
times, and 39 of them are classified as cool core by this criterion.

\subsection{Extracting the SZ profile}\label{extract}

Our primary task is a measurement of the mean ICM pressure profiles by
combining SZ signals from known X-ray clusters.  In ALS05, we assumed a
functional form for the ICM profiles to derive the pressure profile from data.
In this work, we go a step further by developing a model-independent method to
perform these measurements.  This will help minimize systematic uncertainties
due to theoretical modeling of clusters, which are often the dominant sources
of uncertainties in a model-dependent SZ analysis.

The idea is based on the fact that clusters are expected to be
self-similar and exhibit universal dark matter \citep[NFW,][]{nfw}
as well as ICM profiles \citep{univgas,ostriker_etal05}.  Recent
cosmological cluster simulations support that the ICM pressure
profiles in cluster outskirt is remarkably self-similar, even in the
presence of gas cooling and star formation
\citep{daSilva_etal04,motl_etal05,nagai06}.  Since the self-similar
cluster forms a one-parameter family of the ICM pressure profile,
both the radial extent and the amplitude of these profiles can be
scaled by a single parameter, which can be chosen to be the total
mass, X-ray or virial temperature of the cluster.

The key ingredient of our method is to use the observed X-ray temperature of
clusters to set the expected extent of the ICM profile, and use the WMAP maps
to find the mean normalization. We use a simple analytic ICM model to set this
scale: \beq r_{200} = (1.16\Mpc)\left(H(z)\over 100 ~{\rm
    km/s/Mpc}\right)^{-1}\left(T_X\over 5\kev\right)^{1/2}, \label{r200}\eeq
where $r_{200}$ is the radius of the sphere within which the mean density of
the cluster is $200$ times the critical density of the Universe, $\rho_{\rm
  crit}=3H^2(z)/(8\pi G)$, $H(z)$ is the Hubble constant at the redshift of
the cluster, and $T_X$ is the observed X-ray temperature of the
cluster. Following ALS05, the model is based on a spherically
symmetric NFW gravitational potential with concentration
$c_{200}=5$, and a polytropic ICM in hydrostatic equilibrium. Note
that, for a cluster of a given $T_X$, Eq.~\ref{r200} provides
$r_{200}$ that agrees with the best fit scaling relation of the
recent {\it Chandra} observation of X-ray clusters
\citep{vikhlinin_etal06} with the accuracy better than 5$\%$
level\footnote{However, \citet{nagai_etal07} find that the estimated
$r_{200}$ from hydrostatic equilibrium is biased low (by about 10\%)
on average, if subsonic turbulent gas motions are present and their
pressure contribution is not explicitly accounted for \citep[also
see][]{rasia_etal06}. In Sec. \ref{sys_unc}, we discuss the
implications of using a higher normalization for the $r_{200}-T_X$
relation.}.

Once the profiles are scaled at $r_{200}$, we search for the SZ
signal out to $4r_{200}$ radius, since the shock heated gas is only
expected to be present within a few times the virial radius of the
clusters. Given that the CMB signal is correlated on the degree
scale (Doppler peak) in the sky, it is important to distinguish
between primary CMB and SZ signal, and also account for WMAP beam
smearing.  Therefore, we include pixels out to $8\theta_{200}+2
~{\rm deg}$, where $\theta_{200}$ is the angular size of $r_{200}$
in the sky\footnote{For example, for a small cluster, in order to
  find the SZ profile, the mean (large angle) primary CMB background needs to
  be subtracted. Including pixels beyond the cluster helps with setting this
  background.}.  We then aim to constrain spherically averaged pressure
profiles within radial bins that are logarithmically spaced in radius, and are
centered at $0.25$, $0.5$, $1$, $2$, and $4 r_{200}$.  We assume that pressure
is smoothly interpolated as $P_{\rm gas}(r) = A + B r^{-3}$ in between the
centers of the bins, where $A$ and $B$ are constants, and $r^{-3}$ behavior is
motivated by the dark matter/gas/pressure profile in the outskirts of
simulated haloes (see Fig.~\ref{prof}). For completeness, we further assume
$P_{\rm gas}(r) \propto r^{-2}$ within $0.25 r_{200}$ of the cluster center.

The dominant sources of noise for our SZ measurement on angles that can be
resolved by WMAP experiment are WMAP detector noise and CMB primary
anisotropies, which are both expected to be Gaussian to a good approximation.
Therefore, minimizing \beq \chi^2 = \sum_{i,j;a,b} \left[T_{ia}-S_{ia}\right]
C^{-1}_{ia,jb} \left[T_{jb}-S_{jb}\right],\label{chi2} \eeq is the optimum
method of constraining the cluster SZ profile, $S_{ia}$ (which is a
superposition of our radial pressure bins), from the observed CMB temperature
maps $T_{ia}$. Here $i$ and $j$ sum over WMAP pixels, while $a$ and $b$ sum
over WMAP frequency bands, Q(41 GHz), V(61 GHz), W(94 GHz). The noise
covariance matrix $C_{ia,jb}$ is the sum of primary CMB correlation function
\citep[using
CAMB\footnote{\href{http://camb.info/}{http://camb.info/}};][]{2000ApJ...538..473L},
which dominates on large pixel separations, and the WMAP detector noise which
is assumed to be uncorrelated for different pixels/frequencies. Note that both
$C_{ia,jb}$ and $S_{ia}$ need to be convolved with detector beam+pixel window
functions. For WMAP, the effective beam radii range from $0.3~{\rm deg}$ for
Q-band to $0.1~{\rm deg}$ for W band, while we use WMAP foreground cleaned
maps in $N_{side}=512$ HEALPix format (\citealt{gorski_etal05}; $(0.1~ {\rm
  deg})^2$ pixels).  We mask out all the pixels within Kp2 Galactic mask
($\sim 13\%$ of the sky), but do not mask out the point sources, as they are
correlated with the clusters in our sample. In order to deal with a tractable
covariance matrix, we progressively degrade the map pixel resolution towards
the cluster outskirts, so that we have $\lesssim 800$ pixels per cluster.
Moreover, in order to account for the uncertainty in measured $T_X$, as well
as a $15\%$ intrinsic scatter in the $r_{200}-T_X$ relation (seen in our
simulated clusters; see Sec.~\ref{hydro}), we divide the $\chi^2$ for each
cluster by a constant factor.

Point sources are additional sources of contamination of the SZ signal.  The
contamination is expected to be the largest in the lowest frequencies.  So, to
minimize the contamination, we first exclude any cluster that contains a
radial bin with larger than $+3\sigma$ Q-band signal\footnote{Note that the
  systematic bias in the mean due to a $3\sigma$ truncation of a Gaussian
  probability function is less than $0.005\sigma$.}  (remember that SZ signal
is negative for WMAP frequencies). This excludes 18 of our clusters,
only 5 of which have $T_X> 3 \kev$.  Additionally, we assume a radio
point source with a flat spectrum, which is shown to be consistent
with the spectrum of point sources observed by WMAP
\citep{bennett_etal03}, at the center of each cluster. Note,
however, that our results remain virtually unchanged even if we use
a steeper spectrum obtained in \citet{2006astro.ph..8274C} for
fainter cluster radio sources.  To find the ICM pressure profile, we
then marginalize over the mean amplitude of the central source, as
well as a mean point source contamination of the innermost radial
bin, which are expected to be the dominant sources of contamination
of the SZ signal.

\subsection{Hydrodynamic Simulations}\label{hydro}

In this study, we use high-resolution cosmological simulations of nine hot
($T_X>3$ keV), massive galaxy clusters forming in the flat {$\Lambda$}CDM
model: $\Omega_{\rm m}=1-\Omega_{\Lambda}=0.3$, $\Omega_{\rm b}=0.04286$,
$h=0.7$ and $\sigma_8=0.9$, where the Hubble constant is defined as $100h{\
  \rm km\ s^{-1}\ Mpc^{-1}}$, and $\sigma_8$ is the power spectrum
normalization on $8h^{-1}$~Mpc scale.  The simulations follow dissipationless
dark matter and stars, as well as dissipative gas dynamics, and include a
number of physical processes critical to galaxy formation, including radiative
cooling, star formation, stellar feedback and metal enrichment.  The
simulations were performed with the Adaptive Refinement Tree (ART)
$N$-body$+$gas dynamics code \citep{kravtsov99,kravtsov_etal02}, an Eulerian
code that uses adaptive mesh refinement to greatly increase the resolution in
the high density region and resolve formation and evolution of cluster
galaxies and their impact on cluster gas.  Massive clusters with $T_X>3.7$keV,
for example, are simulated in a box size of $120$~Mpc with the spatial
resolution of $3.5\,h^{-1}$kpc and the mass resolution of $9\times
10^8\,h^{-1}\,M_{\odot}$.  Less massive clusters are simulated in a smaller
box of $80$~Mpc in size with higher spatial ($\approx 2.5\,h^{-1}$kpc) and
mass ($3\times 10^{8}\,h^{-1}\,M_{\odot}$) resolutions.  Throughout this work,
we use two different average temperatures of simulated clusters.  The first is
a X-ray spectral temperature, $T_X$, a value derived from a single-temperature
fit to the integrated cluster spectrum within $r_{500}$ extracted from the
mock Chandra images excluding the core ($r<0.15r_{500}$) and detectable
small-scale clumps \citep{nagai_etal07}. We also compute a gas-mass-weighted
temperature, denoted as $\langle T \rangle$ throughout this paper, obtained by
weighting of the 3D temperature with the gas density measured directly in
simulations.  Note that $\langle T \rangle_{200}$ is the gas-mass-weighted
average temperature measured within $r_{200}$.  These simulations have been
used to study the effects of galaxy formation on the Sunyaev-Zel'dovich effect
\citep{nagai06} and the baryon fraction in clusters \citep{kravtsov_etal05}.
These studies are based on a sample of 11 simulated clusters with only 4
clusters massive enough to be included in the current study.  Since then, we
have simulated five more massive clusters, making the total sample of 16
clusters, from which we select nine massive clusters with $T_X>3$~keV for the
current study.  The detailed description and our cluster sample are described
in \citep{nagai_etal07b}.

Note that, in order to compare simulations with observations, the simulated
gas density or pressure is normalized by the ratio of cosmic concordance
baryonic mass fraction \citep[$={\Omega_b}/{\Omega_m}=0.168 \pm
0.007$;][]{spergel06} to the value used in the simulations.

\section{Observational Results}\label{results}
\subsection{Universal Pressure Profile}

\begin{figure}
\includegraphics[width=0.9\linewidth]{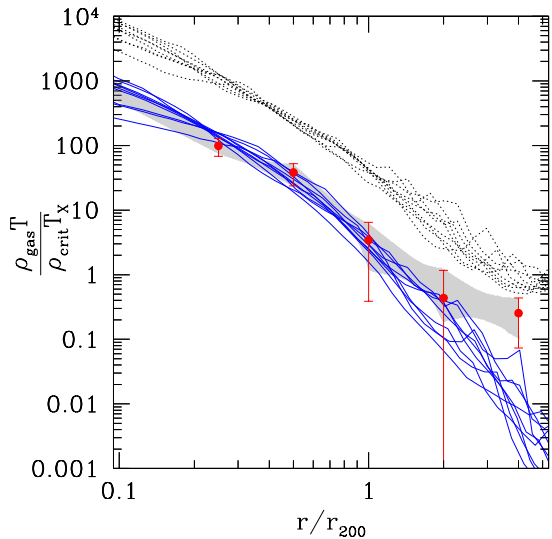}

\caption{Mean pressure profile of 193 of our most massive clusters
  with $T_X > 3\kev$ (points+errorbars). The gray area shows the 68\% region allowed by
this measurement, assuming $P_{\rm gas}>0$ prior, which reflect the
errors, as well as their correlations. The blue/solid curves are
predicted pressure profiles from
  nine simulated clusters with $T_X > 3\kev$, while black/dotted curves show dark
  matter density from the same simulations, divided by the critical density of
  the Universe. \label{prof} }
\end{figure}

The main result of our analysis is shown in Fig.~\ref{prof}.  It
shows the mean ICM pressure profile ($P\equiv \rho_{\rm gas} T$),
normalized to the critical density of the Universe at the cluster's
redshift times its observed X-ray temperature ($\rho_{\rm
crit}T_X$). In this plot, we only include clusters with $T_X >
3\kev$ (or $M_{200} \gtrsim 2.4 \times 10^{14} \msun$). The
points+errorbars show our best fit measurement for a universal
pressure profile.  The gray area shows the 68\% region allowed by
this measurement, assuming $P_{\rm gas}>0$ prior, which reflect the
errors, as well as their correlations. The best fit measurement of
Fig.~\ref{prof} is preferred to null at $\Delta\chi^2 = 47$. The
simulated profiles can improve $\chi^2$ by $31-43$, which implies a
$6-7\sigma$ confirmation of the simulated models, with respect to
null.

Our measurements of the mean ICM pressure profiles closely match the
numerical prediction for the pressure profiles of nine simulated
clusters (indicated by blue/solid lines) in the same temperature
range. It is also interesting to notice the similarity of the
observed ICM pressure and the simulated dark matter density
profiles, indicated by black/dotted lines.

\subsection{Total Thermal Energy of the Cluster}

\begin{figure}
\includegraphics[width=0.9\linewidth]{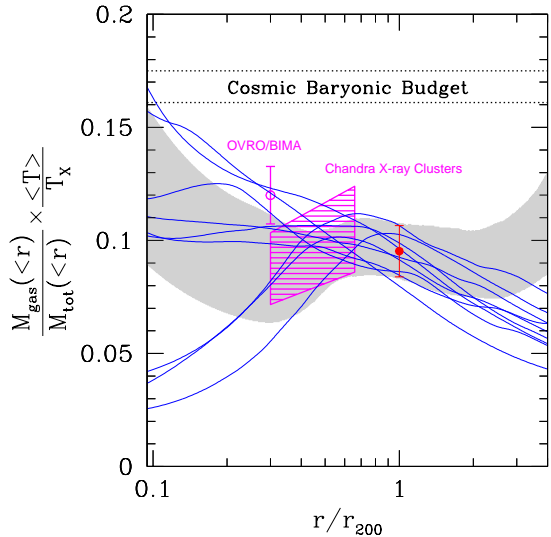}
\caption{The ratio of total thermal energy to the mass times $T_X$,
  for the same clusters as in Fig.~\ref{prof}. The colors are the same as in
  Fig.~\ref{prof}, and we integrate over the pressure profiles to find total
  energy, $M_{\rm gas}\langle T\rangle(<r)$, enclosed within radius $r$, while
  we estimate the total mass from an NFW profile. The dotted lines show the
  total cosmic baryonic budget of $0.168\pm 0.007$ \citep{spergel06}.  The
  dashed region shows the Chandra observational constraints from 8 X-ray
  clusters \citep{vikhlinin_etal06}, while the open point+errorbar is the most
  recent SZ constraint from OVRO/BIMA interferometers \citep{laroque_etal06}.
  \label{energy}}
\end{figure}

In Fig.~(\ref{energy}), we examine the integrated pressure, or equivalently
themal energy of the ICM as a function of cluster centric radius in units of
$r_{200}$.  Here we normalize the thermal energy ($M_{\rm gas,200} \langle
T\rangle_{200}$) to the total mass of the cluster (expected from an NFW
profile with $c_{200}=5$) times the cluster X-ray temperature ($M_{\rm
  tot,200} T_{X}$). The solid point+errorbar shows the mean and
standard deviation of this quantity at $r_{200}$, and thus
corresponds to our measurement of the mean ICM thermal energy:
\beq \left(M_{\rm gas,200}\over M_{\rm tot,200}\right)\left(\langle
  T\rangle_{200}\over T_X\right)_{\rm WMAP}= 0.095 \pm 0.011.
\eeq
The results is in very good agreement with our simulated
clusters:\footnote{The quoted error is the sample variance error of the mean
  energy fraction for the simulated clusters with $T_X > 3\kev$.}
\beq \left(M_{\rm gas,200}\over M_{\rm tot,200}\right)\left(\langle
  T\rangle_{200}\over T_X\right)_{\rm Sim.} = 0.093 \pm 0.003, \eeq
which are shown by thick curves in Fig.~\ref{energy}. Finally, the dashed
region shows the most detailed available X-ray study of 8 clusters
\citep{vikhlinin_etal06} ($T_X > 3\kev$), observed by Chandra X-ray
observatory, while the open dot with the errorbar shows the constraints from
SZ observations of 38 massive clusters with OVRO/BIMA interferometers
\citep{laroque_etal06}.  Both of these studies are also fully consistent with
our SZ measurements from the WMAP 3 year data.

Knowing $\langle T\rangle_{200}/T_X$, we can convert the thermal
energy fraction into the gas fraction of clusters.  For our nine
simulated clusters with $T_X > 3\kev$, we estimate that the ratio of
mean to X-ray estimated temperature is $\langle T\rangle_{200}/T_X =
0.87 \pm 0.02$. Using this value, the cluster gas fraction is
\beq \left({M_{\rm gas,200}}\over{M_{\rm tot,200}} \right)_{\rm
WMAP}=0.109\pm0.013 \eeq
for the WMAP data.  The value is significantly smaller than the
cosmic baryon fraction of $\Omega_{b}/\Omega_{m}=0.168 \pm 0.007$
\citep{spergel06}. If the clusters contain a representative mix of
dark matter and baryons of the Universe ($M_{\rm baryons}/M_{\rm
tot} \approx \Omega_b/\Omega_m$), our results indicate that a
fraction
\beq \frac{M_{\rm missing}}{M_{\rm baryons}}= \frac{M_{\rm
    baryons}-M_{\rm gas}}{M_{\rm baryons}} = 0.35 \pm 0.08\eeq
of the cluster baryons are missing from the hot ICM.  As we discuss
more in \S~\ref{missing}, this number is far too large to be
accounted for by the observed stellar and cold gas fraction of
$\approx 10$\% in our $T_X$ range \citep{gonzalez_etal04,lin04}.
These results are in agreement with our earlier analysis by ALS05.

\subsection{Systematic Trends with $T_X$}\label{systematic}

\begin{figure}
  \includegraphics[width=0.9\linewidth]{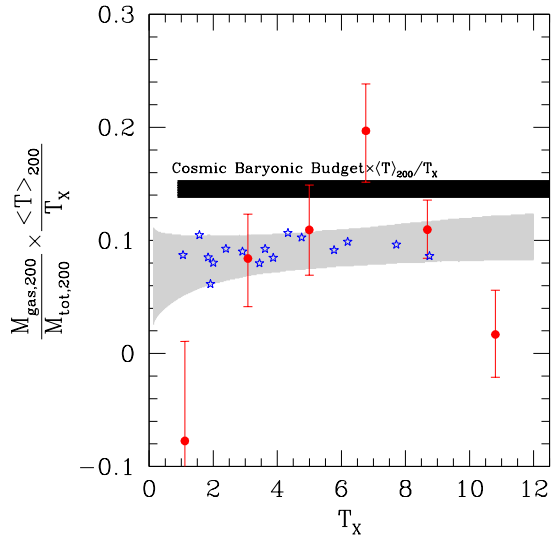}
\caption{ { Thermal energy to mass times $T_X$ ratio as a function
    of $T_X$.} Solid points+errorbars show the measured total thermal energy to mass ratio
  within $r_{200}$ for all our clusters, binned within $\Delta T_X = 2\kev$
  bins.  Open stars show the same ratio for our hydrodynamic simulations, while the gray area
  shows the 68\% uncertainty range of the best power law fit to our data (Eq.~\ref{powerlaw}).
  The black horizontal bar shows the total cosmic baryonic budget
  \citep{spergel06}, multiplied by the mean temperature correction from the simulations.\label{fgastx} }
\end{figure}

Exploiting a large coverage of cluster X-ray temperatures and masses, it would
be interesting to study if there is any systematic trend in the gas fraction
as a function of $T_X$.  Fig.~\ref{fgastx} shows the thermal energy to mass
ratio within $r_{200}$ as a function $T_X$, where we have binned clusters into
$2\kev$ temperature bins.  The number of clusters in each bin, from left to
right is 44, 66, 84, 41, 16, and 9. Here again, we see that our SZ
measurements are consistent with predictions from hydrodynamic simulations
(open stars), as well as our earlier analysis (ALS05). The best fit power law
to our measurement points (weighted by their errors) is \beq \left(M_{\rm
    gas,200}\over M_{\rm tot,200}\right)\left(\langle T\rangle_{200}\over
  T_X\right) = (0.098 \pm 0.015) \left(T_X\over 6.4\kev \right)^{0.15 \pm
  0.18}, \label{powerlaw}\eeq which is represented by the gray area in
Fig.~\ref{fgastx}, and does not show any significant monotonic trend
throughout the $T_X$ range, a trend similar to the simulated clusters.

The measurement points in Fig.~\ref{fgastx} have almost a factor of two more
scatter than is expected from the measurement errors.  This could be caused by
a possible systematic underestimate of our measurement errors, or otherwise by
an intrinsic scatter in the ICM properties of different clusters.  An
important clue to the reason behind this scatter, as well as the nature of
missing baryons may come from looking at a sub-sample of our clusters with
resolved cool cores (39 CC-clusters with $T_X > 3\kev$; see Sec.
\ref{x-ray}).  The energy fraction of the CC-clusters is \beq \left(M_{\rm
    gas,200}\over M_{\rm tot,200}\right)\left(\langle T\rangle_{200}\over
  T_X\right)_{\rm CC} = 0.142 \pm 0.019,\label{cool}\eeq which is
significantly larger than the main sample, and is consistent with no missing
baryons after the correction for $\langle T\rangle_{200}/T_X$.
Moreover, the pressure profile seems significantly flatter than the simulated
profiles within $0.5 r_{200}$. Although the fact that the ICM gas fraction is
systematically higher for cool core clusters in Eq.~\ref{cool} may signal an
unknown systematic problem with our method of subtracting point sources, it
may also be caused by a real anti-correlation between the missing baryons and
the central cooling activity.  Further observational studies are clearly
needed to resolve these issues.

Although our sample includes X-ray clusters with measured
temperatures out to redshift 1, less than 20\% of the statistical
significance of our SZ detection in our sample comes from clusters
with $z > 0.2$. This is because the virial radius of clusters ($\sim
1.5 \Mpc$) at $z\gtrsim 0.2$ falls below the WMAP detector beam size
of $\sim 0.2 ~{\rm deg}$. Therefore, it would be difficult to study
any redshift evolution in our sample, and thus it will not be
explored here.

\section{Discussions}\label{discuss}

\subsection{Comparison with our previous analysis}

One may wonder why, despite the use of the 3 year WMAP maps as well
as more X-ray clusters, the significance of our SZ detection has not
significantly increased compared to our previous analysis with the
1st year WMAP maps in ALS05. This is mainly due to the fact that, in
our new analysis we parameterize the mean cluster signal by 8 free
parameters (6 for pressure profile, and 2 for point sources), while
our previous analysis only used 2 (1 for the total gas fraction, and
1 for the over all point source contamination).  Therefore, the
error on the individual parameters, such as the total gas fraction,
is increased due to the extra marginalizations involved in the
analysis.

Despite the use of a completely different method, our main
conclusion: a fraction of $30\%-40\%$ missing ICM baryons, has not
changed from ALS05. However, we did not find any systematic $T_X$
dependence in the energy fraction in Eq.~\ref{powerlaw}. This shows
a $\sim 2\sigma$ discrepancy with our earlier analysis of WMAP SZ
signal, where we found an almost linear dependence on $T_X$.
However, our earlier analysis assumed a universal pressure profile
for all clusters. Therefore, one way to reconcile the two analyses
is for the profiles to be systematically shallower for clusters with
$T_X \lesssim 4 \kev$, which is indeed seen in our sample, but it is
difficult to quantify due to the low significance of SZ signal in
cooler clusters.

\subsection{Systematic and Theoretical Uncertainties}\label{sys_unc}

One of the main improvements from our previous analysis in ALS05 is
the use of a model-independent method to derive the mean ICM
profiles from the WMAP 3yr data.  This will help minimize systematic
errors arising from uncertainties in theoretical modeling, which
often dominate systematic uncertainties of the analyses carried out
using a model-dependent method.

The only theoretical assumption in our derivation of the mean ICM
pressure profile has been the use of Eq.~\ref{r200} to infer the
cluster masses or radii based on their measured X-ray temperatures.
Therefore, any systematic errors in this scaling relation translate
into an error in the cluster total mass estimate, which could bias
our estimate of the cluster gas mass fractions. Indeed, the
normalization of the $r_{200}-T_X$ used in our analysis is
consistent with hydrostatic estimates from X-ray observations
\citep{vikhlinin_etal06}, which are about 10\% smaller than the
simulated normalization \citep{kravtsov_etal06}.
However, even if there is a systematic error in the normalization of
$r_{200}-T_X$ relation, Eq.~\ref{r200} can be viewed as the
definition of $r_{200}$ in terms of the observed cluster X-ray
temperature, which simply defines a unified radial unit for the mean
pressure profile. In other words, the bias, if present, results in
the re-definition (or re-scaling) of $r_{200}$. In this way, our
measurements of mean pressure profile is indeed {\it
model-independent}.  However, any inference about the missing ICM
baryonic fraction would be affected by a systematic error in our
assumed $r_{200}-T_X$ relation. For example, using the higher
normalization (inferred from simulations) can boost the missing
baryon fraction to $\sim 50\%$.

\subsection{Missing Baryons or New Astrophysics?}\label{missing}

The most puzzling result of our analysis is that a significant
fraction ($35 \%\pm 8 \%$) of the baryons are missing from the ICM.
On average, the fraction of the observed stellar mass associated
with galaxies in clusters within our $T_X$ range is less than $10\%$
\citep{lin_etal03}, leaving room for an un-accounted baryonic
component of $25 \pm 8 \%$. Although the evidence for this component
is marginal ($3.1\sigma$) from our analysis, independent
verification from other SZ and X-ray observations \citep[][see
Fig.~\ref{energy}]{laroque_etal06,ettori_etal03,vikhlinin_etal06}
strongly support the case for a missing component of cluster
baryons.

Interestingly, modern cosmological cluster simulations that include
gas cooling and star formation also give the ICM mass fraction very
similar to the observed values \citep{kravtsov_etal07}.  The good
agreement in the ICM mass fraction between simulations and
observations suggests that the simulations reproduce the hot ICM of
the real clusters reasonably well \citep {nagai_etal07}. However,
the agreement is achieved while the stellar fractions within
$r_{500}$ of the simulated clusters are a factor of two to three
greater than the observed stellar fraction.

This raises a possibility that a significant fraction of baryonic
mass may be hidden in a form that is difficult to detect with
ordinary observational means.  One such example is diffuse
intracluster light (ICL), which would be challenging to detect
observationally, but could potentially hide significant amount of
stellar mass in the vast intracluster space.  The measurements of
ICL are now becoming possible with very deep photometry. Although
the measurements are quite challenging, a number of ICL observations
indicate that no more than 10-50\% of stars could be found in the
form of ICL
\citep{lin04,zibetti_etal05,gonzalez_etal04,monaco_etal06}. Similar
results are also obtained in cosmological cluster simulations
\citep{murante04,larsen_etal05}.  Alternatively, one may also
envisage hiding baryons in a form of cold and compact dark baryonic
clouds formed from local cooling instabilities within ICM (or
proto-ICM), similar to the high velocity clouds in the local group.
At the moment, there is essentially no constraint on such clouds as
long as their temperature is below 10~K, and/or their covering
fraction of the cluster radio map is less than unity
\citep{dwarakanath_etal94}.

It has been suggested that a soft X-ray excess in observations of
X-ray clusters can be interpreted as evidence for a warm ($\sim 0.1
\kev$) compact phase of the ICM, which would not contribute to the
SZ signal, while containing a significant fraction of the ICM
baryons \citep[\eg][and references therein]{2005ApJ...629..192B}.
However, the extragalactic nature of the soft X-ray excess is still
controversial \citep[\eg][]{2006ApJ...644..167B}. Moreover, it is
not clear how the strong cooling instability of any such component
could be avoided in the cluster environment.

Yet another possibility that has been recently proposed by
\citet{loeb06}, is thermal diffusion or evaporation of baryons out
of the virial radius of the cluster. The thermal evaporation, if
efficient, can remove the gas out of the virial radius, bringing the
cluster gas fraction down without creating stars \citep[but
see][]{2007astro.ph..2356M}. Similarly, inclusion of diffusive
processes in hydrodynamical simulations may on one hand suppress
cooling \citep{zakamska_etal03}, thus bringing down the cool gas
fraction into better agreement with the observed stellar content,
while on the other hand remove the thermal energy of the ICM out of
the virial radius \citep{loeb_etal02}, explaining the low gas
fractions inferred from our SZ observations.  However, recent
numerical simulations that included (some of) these processes
suggest that they cannot affect the overcooling problem
significantly \citep{dolag_etal04}.


%
%


\section{Summary and Future Prospects} \label{sec:future}

We report a $9\sigma$ detection of the SZ signal in the CMB maps of WMAP 3yr
data, through study of a sample of 193 massive clusters with observed X-ray
temperatures larger than $3\kev$.  The improvements from the previous analyses
in ALS05 are in both data and analysis methods: (1) the use of the WMAP 3yr
data release \citep{hinshaw06} and (2) the use of a model-independent method
to measure the mean ICM pressure profile to minimize systematic uncertainties
due to theoretical modeling.

The resulting pressure profile is in good agreement with
measurements of recent X-ray and SZ observations as well as
hydrodynamical cluster simulations.  Our result indicates that
$35\pm 8\%$ of the baryons are missing from the ICM, which is
significantly larger than the observed stellar or cold gas fraction
of about 10\% so far accounted for in clusters.  The evidence for
this component is marginal ($3.1\sigma$) from our analysis, but
independent verification from other SZ \citep{laroque_etal06} and
X-ray observations \citep{ettori_etal03,vikhlinin_etal06} strongly
support the case for a missing component of cluster baryons.  This
signals the presence of a missing baryonic component or yet-unknown
astrophysical processes that could lower the cluster baryon
fraction.

The problem of missing thermal energy of the ICM introduces a new
puzzle into the standard cluster formation scenario.  Further
investigations of the ICM physics are thus essential for our
understanding of cluster formation as well as applications of
cluster-based cosmological tests using X-ray and SZ cluster surveys.
On the theoretical front, it is important to understand details and
relative importance of gas cooling and heating by supernovae (SN)
and active galactic nuclei (AGN) and their effects on the ICM
properties.  Further numerical investigations of diffusive processes
are also critical for understanding of cluster plasma phenomena.
Observationally, it is important to further refine measurements of
X-ray, SZ, lensing and optical/IR observations and consolidate the
measurements of the cluster baryon budgets in clusters. Large
microwave interferometer arrays such as
ALMA\footnote{\href{http://www.alma.nrao.edu/}{http://www.alma.nrao.edu/}}
will ultimately be able to map the ICM pressure profile in the
outskirts of individual clusters with exquisite accuracy
\citep{2005ApJ...623..632K}, while the next generation of
high-resolution, multi-frequency CMB experiments, such as APEX, ACT,
SPT, and
SZA\footnote{\href{http://bolo.berkeley.edu/apexsz/}{http://bolo.berkeley.edu/apexsz/};
  \href{http://www.physics.princeton.edu/act/}{http://www.physics.princeton.edu/act/};
  \href{http://spt.uchicago.edu/}{http://spt.uchicago.edu/};
  \href{http://astro.uchicago.edu/sza/}{http://astro.uchicago.edu/sza/}}, will
be able to probe the thermal energy content to high redshifts,
opening new windows for the future studies of structure formation
and cosmology.



\section*{Acknowledgments}
We would like to thank Jacqueline van Gorkom, Andrey Kravtsov, Adam
Lidz, Avi Loeb, Ramesh Narayan, and Jerry Ostriker for helpful
comments and discussions. YTL acknowledges support from NSF PIRE
grant OISE-0530095 and FONDAP-Andes. DN is supported by the Sherman
Fairchild Postdoctoral Fellowship at CalTech. AJRS acknowledges
support by PPARC.

The analysis presented in this paper have made use of the HEALPix
package
\citep[\href{http://healpix.jpl.nasa.gov/}{http://healpix.jpl.nasa.gov/}]{gorski_etal05},
the X-Ray Galaxy Clusters Database (BAX;
\href{http://bax.ast.obs-mip.fr/}{http://bax.ast.obs-mip.fr/}), and
the Legacy Archive for Microwave Background Data Analysis (LAMBDA;
\href{http://lambda.gsfc.nasa.gov/}{http://lambda.gsfc.nasa.gov/}).

\bibliography{szmap2}

\begin{thebibliography}{}

\bibitem[\protect\citeauthoryear{{Afshordi}, {Lin} \& {Sanderson}}{{Afshordi}
  et~al.}{2005}]{ALS}
{Afshordi} N.,  {Lin} Y.-T.,    {Sanderson} A.~J.~R.,  2005, ApJ, 629, 1,
  \adsurl{http://adsabs.harvard.edu/cgi-bin/nph-bib_query?bibcode=2005ApJ...62%
9....1A&db_key=AST}, \eprint{astro-ph/0408560}

\bibitem[\protect\citeauthoryear{{Afshordi}, {Loh} \& {Strauss}}{{Afshordi}
  et~al.}{2004}]{afshordi_etal04}
{Afshordi} N.,  {Loh} Y.-S.,    {Strauss} M.~A.,  2004, \prd, 69, 083524,
  \adsurl{http://adsabs.harvard.edu/cgi-bin/nph-bib_query?bibcode=2004PhRvD..6%
9h3524A&db_key=AST}, \eprint{astro-ph/0308260}

\bibitem[\protect\citeauthoryear{{Bauer}, {Fabian}, {Sanders}, {Allen} \&
  {Johnstone}}{{Bauer} et~al.}{2005}]{bauer05}
{Bauer} F.~E.,  {Fabian} A.~C.,  {Sanders} J.~S.,  {Allen} S.~W.,
  {Johnstone} R.~M.,  2005, MNRAS, 359, 1481,
  \adsurl{http://adsabs.harvard.edu/cgi-bin/nph-bib_query?bibcode=2005MNRAS.35%
9.1481B&db_key=AST}, \eprint{astro-ph/0503232}

\bibitem[\protect\citeauthoryear{{Bennett} et~al.,}{{Bennett}
  et~al.}{2003a}]{bennett_etal03}
{Bennett} C.~L.,  et~al., 2003a, \apjs, 148, 97, \eprint{astro-ph/0302208}

\bibitem[\protect\citeauthoryear{{Bennett} et~al.,}{{Bennett}
  et~al.}{2003b}]{bennett03}
{Bennett} C.~L.,  et~al., 2003b, \apjs, 148, 1,
  \adsurl{http://adsabs.harvard.edu/cgi-bin/nph-bib_query?bibcode=2003ApJS..14%
8....1B&db_key=AST}, \eprint{astro-ph/0302207}

\bibitem[\protect\citeauthoryear{{Bialek}, {Evrard} \& {Mohr}}{{Bialek}
  et~al.}{2001}]{bialek_etal01}
{Bialek} J.~J.,  {Evrard} A.~E.,    {Mohr} J.~J.,  2001, ApJ, 555, 597,
  \adsurl{http://adsabs.harvard.edu/cgi-bin/nph-bib_query?bibcode=2001ApJ...55%
5..597B&db_key=AST}, \eprint{astro-ph/0010584}

\bibitem[\protect\citeauthoryear{{Birkinshaw}}{{Birkinshaw}}{1999}]{birkinshaw%
_etal99}
{Birkinshaw} M.,  1999, \physrep, 310, 97,
  \adsurl{http://adsabs.harvard.edu/cgi-bin/nph-bib_query?bibcode=1999PhR...31%
0...97B&db_key=AST}, \eprint{astro-ph/9808050}

\bibitem[\protect\citeauthoryear{{Bonamente}, {Lieu}, {Mittaz}, {Kaastra} \&
  {Nevalainen}}{{Bonamente} et~al.}{2005}]{2005ApJ...629..192B}
{Bonamente} M.,  {Lieu} R.,  {Mittaz} J.~P.~D.,  {Kaastra} J.~S.,
  {Nevalainen} J.,  2005, ApJ, 629, 192,
  \adsurl{http://adsabs.harvard.edu/cgi-bin/nph-bib_query?bibcode=2005ApJ...62%
9..192B&db_key=AST}, \eprint{astro-ph/0504092}

\bibitem[\protect\citeauthoryear{{Borgani}}{{Borgani}}{2006}]{borgani_etal06}
{Borgani} S.,  2006, ArXiv Astrophysics e-prints, \eprint{astro-ph/0605575}

\bibitem[\protect\citeauthoryear{{Bregman} \& {Lloyd-Davies}}{{Bregman} \&
  {Lloyd-Davies}}{2006}]{2006ApJ...644..167B}
{Bregman} J.~N.,  {Lloyd-Davies} E.~J.,  2006, ApJ, 644, 167,
  \adsurl{http://adsabs.harvard.edu/cgi-bin/nph-bib_query?bibcode=2006ApJ...64%
4..167B&db_key=AST}, \eprint{astro-ph/0602527}

\bibitem[\protect\citeauthoryear{{Carlstrom}, {Holder} \& {Reese}}{{Carlstrom}
  et~al.}{2002}]{carlstrom_etal02}
{Carlstrom} J.~E.,  {Holder} G.~P.,    {Reese} E.~D.,  2002, \araa, 40, 643,
  \adsurl{http://adsabs.harvard.edu/cgi-bin/nph-bib_query?bibcode=2002ARA
.40..643C&db_key=AST}, \eprint{astro-ph/0208192}

\bibitem[\protect\citeauthoryear{{Coble}, {Carlstrom}, {Bonamente}, {Dawson},
  {Holzapfel}, {Joy}, {LaRoque} \& {Reese}}{{Coble}
  et~al.}{2006}]{2006astro.ph..8274C}
{Coble} K.,  {Carlstrom} J.~E.,  {Bonamente} M.,  {Dawson} K.,  {Holzapfel} W.,
   {Joy} M.,  {LaRoque} S.,    {Reese} E.~D.,  2006, ArXiv Astrophysics
  e-prints, \eprint{astro-ph/0608274}

\bibitem[\protect\citeauthoryear{{da Silva}, {Kay}, {Liddle} \& {Thomas}}{{da
  Silva} et~al.}{2004}]{daSilva_etal04}
{da Silva} A.~C.,  {Kay} S.~T.,  {Liddle} A.~R.,    {Thomas} P.~A.,  2004,
  MNRAS, 348, 1401, \eprint{astro-ph/0308074}

\bibitem[\protect\citeauthoryear{{Dolag}, {Jubelgas}, {Springel}, {Borgani} \&
  {Rasia}}{{Dolag} et~al.}{2004}]{dolag_etal04}
{Dolag} K.,  {Jubelgas} M.,  {Springel} V.,  {Borgani} S.,    {Rasia} E.,
  2004, ApJ, 606, L97, \eprint{astro-ph/0401470}

\bibitem[\protect\citeauthoryear{{Dwarakanath}, {van Gorkom} \&
  {Owen}}{{Dwarakanath} et~al.}{1994}]{dwarakanath_etal94}
{Dwarakanath} K.~S.,  {van Gorkom} J.~H.,    {Owen} F.~N.,  1994, ApJ, 432,
  469,
  \adsurl{http://adsabs.harvard.edu/cgi-bin/nph-bib_query?bibcode=1994ApJ...43%
2..469D&db_key=AST}

\bibitem[\protect\citeauthoryear{{Eke}, {Navarro} \& {Frenk}}{{Eke}
  et~al.}{1998}]{eke_etal98}
{Eke} V.~R.,  {Navarro} J.~F.,    {Frenk} C.~S.,  1998, ApJ, 503, 569,
  \adsurl{http://adsabs.harvard.edu/cgi-bin/nph-bib_query?bibcode=1998ApJ...50%
3..569E&db_key=AST}, \eprint{astro-ph/9708070}

\bibitem[\protect\citeauthoryear{{Ettori}}{{Ettori}}{2003}]{ettori_etal03}
{Ettori} S.,  2003, MNRAS, 344, L13,
  \adsurl{http://adsabs.harvard.edu/cgi-bin/nph-bib_query?bibcode=2003MNRAS.34%
4L..13E&db_key=AST}, \eprint{astro-ph/0305296}

\bibitem[\protect\citeauthoryear{{Ettori}, {Dolag}, {Borgani} \&
  {Murante}}{{Ettori} et~al.}{2006}]{ettori_etal06}
{Ettori} S.,  {Dolag} K.,  {Borgani} S.,    {Murante} G.,  2006, MNRAS, 365,
  1021, \eprint{astro-ph/0509024}

\bibitem[\protect\citeauthoryear{{Evrard}}{{Evrard}}{1990}]{evrard90}
{Evrard} A.~E.,  1990, ApJ, 363, 349,
  \adsurl{http://adsabs.harvard.edu/cgi-bin/nph-bib_query?bibcode=1990ApJ...36%
3..349E&db_key=AST}

\bibitem[\protect\citeauthoryear{{Finoguenov}, {Reiprich} \&
  {B{\"o}hringer}}{{Finoguenov} et~al.}{2001}]{finoguenov01}
{Finoguenov} A.,  {Reiprich} T.~H.,    {B{\"o}hringer} H.,  2001, A\&A, 368,
  749, \eprint{astro-ph/0010190}

\bibitem[\protect\citeauthoryear{{Fosalba} \& {Gazta{\~n}aga}}{{Fosalba} \&
  {Gazta{\~n}aga}}{2004}]{fosalba_etal04}
{Fosalba} P.,  {Gazta{\~n}aga} E.,  2004, MNRAS, 350, L37,
  \eprint{astro-ph/0305468}

\bibitem[\protect\citeauthoryear{{Fosalba}, {Gazta{\~n}aga} \&
  {Castander}}{{Fosalba} et~al.}{2003}]{fosalba_etal03}
{Fosalba} P.,  {Gazta{\~n}aga} E.,    {Castander} F.~J.,  2003, ApJ, 597, L89,
  \eprint{astro-ph/0307249}

\bibitem[\protect\citeauthoryear{{Frenk} et~al.,}{{Frenk}
  et~al.}{1999}]{frenk_etal99}
{Frenk} C.~S.,  et~al., 1999, ApJ, 525, 554, \eprint{astro-ph/9906160}

\bibitem[\protect\citeauthoryear{{Fukazawa}, {Makishima} \&
  {Ohashi}}{{Fukazawa} et~al.}{2004}]{fukazawa04}
{Fukazawa} Y.,  {Makishima} K.,    {Ohashi} T.,  2004, \pasj, 56, 965,
  \eprint{astro-ph/0411745}

\bibitem[\protect\citeauthoryear{{Geisbuesch} \& {Hobson}}{{Geisbuesch} \&
  {Hobson}}{2006}]{2006astro.ph.11567G}
{Geisbuesch} J.,  {Hobson} M.,  2006, ArXiv Astrophysics e-prints,
  \adsurl{http://adsabs.harvard.edu/cgi-bin/nph-bib_query?bibcode=2006astro.ph%
.11567G&db_key=PRE}, \eprint{astro-ph/0611567}

\bibitem[\protect\citeauthoryear{{Gonzalez}, {Zabludoff} \&
  {Zaritsky}}{{Gonzalez} et~al.}{2005}]{gonzalez_etal04}
{Gonzalez} A.~H.,  {Zabludoff} A.~I.,    {Zaritsky} D.,  2005, ApJ, 618, 195,
  \adsurl{http://adsabs.harvard.edu/cgi-bin/nph-bib_query?bibcode=2005ApJ...61%
8..195G&db_key=AST}, \eprint{astro-ph/0406244}

\bibitem[\protect\citeauthoryear{{G{\'o}rski}, {Hivon}, {Banday}, {Wandelt},
  {Hansen}, {Reinecke} \& {Bartelmann}}{{G{\'o}rski}
  et~al.}{2005}]{gorski_etal05}
{G{\'o}rski} K.~M.,  {Hivon} E.,  {Banday} A.~J.,  {Wandelt} B.~D.,  {Hansen}
  F.~K.,  {Reinecke} M.,    {Bartelmann} M.,  2005, ApJ, 622, 759,
  \eprint{astro-ph/0409513}

\bibitem[\protect\citeauthoryear{{Hern{\'a}ndez-Monteagudo} \&
  {Rubi{\~n}o-Mart{\'{\i}}n}}{{Hern{\'a}ndez-Monteagudo} \&
  {Rubi{\~n}o-Mart{\'{\i}}n}}{2004}]{hernandez_etal04}
{Hern{\'a}ndez-Monteagudo} C.,  {Rubi{\~n}o-Mart{\'{\i}}n} J.~A.,  2004, MNRAS,
  347, 403, \eprint{astro-ph/0305606}

\bibitem[\protect\citeauthoryear{{Hinshaw} et~al.,}{{Hinshaw}
  et~al.}{2006}]{hinshaw06}
{Hinshaw} G.,  et~al., 2006, ArXiv Astrophysics e-prints,
  \adsurl{http://adsabs.harvard.edu/cgi-bin/nph-bib_query?bibcode=2006astro.ph%
..3451H&db_key=PRE}, \eprint{astro-ph/0603451}

\bibitem[\protect\citeauthoryear{{Ikebe}, {Reiprich}, {B{\"o}hringer}, {Tanaka}
  \& {Kitayama}}{{Ikebe} et~al.}{2002}]{ikebe02}
{Ikebe} Y.,  {Reiprich} T.~H.,  {B{\"o}hringer} H.,  {Tanaka} Y.,    {Kitayama}
  T.,  2002, A\&A, 383, 773, \eprint{astro-ph/0112315}

\bibitem[\protect\citeauthoryear{{Jarosik} et~al.,}{{Jarosik}
  et~al.}{2006}]{2006astro.ph..3452J}
{Jarosik} N.,  et~al., 2006, ArXiv Astrophysics e-prints,
  \adsurl{http://adsabs.harvard.edu/cgi-bin/nph-bib_query?bibcode=2006astro.ph%
..3452J&db_key=PRE}, \eprint{astro-ph/0603452}

\bibitem[\protect\citeauthoryear{{Kocsis}, {Haiman} \& {Frei}}{{Kocsis}
  et~al.}{2005}]{2005ApJ...623..632K}
{Kocsis} B.,  {Haiman} Z.,    {Frei} Z.,  2005, ApJ, 623, 632,
  \adsurl{http://adsabs.harvard.edu/cgi-bin/nph-bib_query?bibcode=2005ApJ...62%
3..632K&db_key=AST}, \eprint{astro-ph/0409430}

\bibitem[\protect\citeauthoryear{{Komatsu} \& {Seljak}}{{Komatsu} \&
  {Seljak}}{2001}]{univgas}
{Komatsu} E.,  {Seljak} U.,  2001, MNRAS, 327, 1353,
  \adsurl{http://adsabs.harvard.edu/cgi-bin/nph-bib_query?bibcode=2001MNRAS.32%
7.1353K&db_key=AST}, \eprint{astro-ph/0106151}

\bibitem[\protect\citeauthoryear{{Kravtsov}}{{Kravtsov}}{1999}]{kravtsov99}
{Kravtsov} A.~V.,  1999, PhD thesis, New Mexico State University

\bibitem[\protect\citeauthoryear{{Kravtsov}, {Klypin} \& {Hoffman}}{{Kravtsov}
  et~al.}{2002}]{kravtsov_etal02}
{Kravtsov} A.~V.,  {Klypin} A.,    {Hoffman} Y.,  2002, ApJ, 571, 563,
  \eprint{astro-ph/0109077}

\bibitem[\protect\citeauthoryear{{Kravtsov}, {Nagai} \& {Vikhlinin}}{{Kravtsov}
  et~al.}{2005}]{kravtsov_etal05}
{Kravtsov} A.~V.,  {Nagai} D.,    {Vikhlinin} A.~A.,  2005, ApJ, 625, 588,
  \eprint{astro-ph/0501227}

\bibitem[\protect\citeauthoryear{{Kravtsov}, {Nagai} \& {Vikhlinin}}{{Kravtsov}
  et~al.}{2007}]{kravtsov_etal07}
{Kravtsov} A.~V.,  {Nagai} D.,    {Vikhlinin} A.~A.,  2007, \apj, in
  preparation

\bibitem[\protect\citeauthoryear{{Kravtsov}, {Vikhlinin} \& {Nagai}}{{Kravtsov}
  et~al.}{2006}]{kravtsov_etal06}
{Kravtsov} A.~V.,  {Vikhlinin} A.,    {Nagai} D.,  2006, ApJ, 650, 128,
  \adsurl{http://adsabs.harvard.edu/cgi-bin/nph-bib_query?bibcode=2006ApJ...65%
0..128K&db_key=AST}, \eprint{astro-ph/0603205}

\bibitem[\protect\citeauthoryear{{LaRoque}, {Bonamente}, {Carlstrom}, {Joy},
  {Nagai}, {Reese} \& {Dawson}}{{LaRoque} et~al.}{2006}]{laroque_etal06}
{LaRoque} S.~J.,  {Bonamente} M.,  {Carlstrom} J.~E.,  {Joy} M.~K.,  {Nagai}
  D.,  {Reese} E.~D.,    {Dawson} K.~S.,  2006, ApJ, 652, 917,
  \adsurl{http://adsabs.harvard.edu/cgi-bin/nph-bib_query?bibcode=2006ApJ...65%
2..917L&db_key=AST}, \eprint{astro-ph/0604039}

\bibitem[\protect\citeauthoryear{{Lewis}, {Challinor} \& {Lasenby}}{{Lewis}
  et~al.}{2000}]{2000ApJ...538..473L}
{Lewis} A.,  {Challinor} A.,    {Lasenby} A.,  2000, ApJ, 538, 473,
  \eprint{astro-ph/9911177}

\bibitem[\protect\citeauthoryear{{Lieu}, {Mittaz} \& {Zhang}}{{Lieu}
  et~al.}{2006}]{2006ApJ...648..176L}
{Lieu} R.,  {Mittaz} J.~P.~D.,    {Zhang} S.-N.,  2006, ApJ, 648, 176,
  \adsurl{http://adsabs.harvard.edu/cgi-bin/nph-bib_query?bibcode=2006ApJ...64%
8..176L&db_key=AST}, \eprint{astro-ph/0510160}

\bibitem[\protect\citeauthoryear{{Lin} \& {Mohr}}{{Lin} \&
  {Mohr}}{2004}]{lin04}
{Lin} Y.-T.,  {Mohr} J.~J.,  2004, ApJ, 617, 879, \eprint{astro-ph/0408557}

\bibitem[\protect\citeauthoryear{{Lin}, {Mohr} \& {Stanford}}{{Lin}
  et~al.}{2003}]{lin_etal03}
{Lin} Y.-T.,  {Mohr} J.~J.,    {Stanford} S.~A.,  2003, ApJ, 591, 749,
  \eprint{astro-ph/0304033}

\bibitem[\protect\citeauthoryear{{Loeb}}{{Loeb}}{2002}]{loeb_etal02}
{Loeb} A.,  2002, New Astronomy, 7, 279, \eprint{astro-ph/0203450}

\bibitem[\protect\citeauthoryear{{Loeb}}{{Loeb}}{2007}]{loeb06}
{Loeb} A.,  2007, Journal of Cosmology and Astro-Particle Physics, 3, 1,
  \adsurl{http://adsabs.harvard.edu/cgi-bin/nph-bib_query?bibcode=2007JCAP...0%
3....1L&db_key=AST}, \eprint{astro-ph/0606572}

\bibitem[\protect\citeauthoryear{{Lubin}, {Cen}, {Bahcall} \&
  {Ostriker}}{{Lubin} et~al.}{1996}]{lubin_etal96}
{Lubin} L.~M.,  {Cen} R.,  {Bahcall} N.~A.,    {Ostriker} J.~P.,  1996, ApJ,
  460, 10,
  \adsurl{http://adsabs.harvard.edu/cgi-bin/nph-bib_query?bibcode=1996ApJ...46%
0...10L&db_key=AST}, \eprint{astro-ph/9509148}

\bibitem[\protect\citeauthoryear{{McCarthy}, {Bower} \& {Balogh}}{{McCarthy}
  et~al.}{2006}]{mccarthy_etal06}
{McCarthy} I.~G.,  {Bower} R.~G.,    {Balogh} M.~L.,  2006, ArXiv Astrophysics
  e-prints, \eprint{astro-ph/0609314}

\bibitem[\protect\citeauthoryear{{Medvedev}}{{Medvedev}}{2007}]{2007astro.ph..%
2356M}
{Medvedev} M.~V.,  2007, ArXiv Astrophysics e-prints,
  \adsurl{http://adsabs.harvard.edu/cgi-bin/nph-bib_query?bibcode=2007astro.ph%
..2356M&db_key=PRE}, \eprint{astro-ph/0702356}

\bibitem[\protect\citeauthoryear{{Metzler} \& {Evrard}}{{Metzler} \&
  {Evrard}}{1994}]{metzler_evrard94}
{Metzler} C.~A.,  {Evrard} A.~E.,  1994, ApJ, 437, 564,
  \adsurl{http://adsabs.harvard.edu/cgi-bin/nph-bib_query?bibcode=1994ApJ...43%
7..564M&db_key=AST}, \eprint{astro-ph/9309050}

\bibitem[\protect\citeauthoryear{{Mohr}, {Mathiesen} \& {Evrard}}{{Mohr}
  et~al.}{1999}]{mohr_etal99}
{Mohr} J.~J.,  {Mathiesen} B.,    {Evrard} A.~E.,  1999, ApJ, 517, 627,
  \adsurl{http://adsabs.harvard.edu/cgi-bin/nph-bib_query?bibcode=1999ApJ...51%
7..627M&db_key=AST}, \eprint{astro-ph/9901281}

\bibitem[\protect\citeauthoryear{{Monaco}, {Murante}, {Borgani} \&
  {Fontanot}}{{Monaco} et~al.}{2006}]{monaco_etal06}
{Monaco} P.,  {Murante} G.,  {Borgani} S.,    {Fontanot} F.,  2006, ApJ, 652,
  L89, \eprint{astro-ph/0610045}

\bibitem[\protect\citeauthoryear{{Motl}, {Hallman}, {Burns} \& {Norman}}{{Motl}
  et~al.}{2005}]{motl_etal05}
{Motl} P.~M.,  {Hallman} E.~J.,  {Burns} J.~O.,    {Norman} M.~L.,  2005, ApJ,
  623, L63, \eprint{astro-ph/0502226}

\bibitem[\protect\citeauthoryear{{Murante}, {Arnaboldi}, {Gerhard}, {Borgani},
  {Cheng}, {Diaferio}, {Dolag}, {Moscardini}, {Tormen}, {Tornatore} \&
  {Tozzi}}{{Murante} et~al.}{2004}]{murante04}
{Murante} G.,  {Arnaboldi} M.,  {Gerhard} O.,  {Borgani} S.,  {Cheng} L.~M.,
  {Diaferio} A.,  {Dolag} K.,  {Moscardini} L.,  {Tormen} G.,  {Tornatore} L.,
    {Tozzi} P.,  2004, ApJ, 607, L83, \eprint{astro-ph/0404025}

\bibitem[\protect\citeauthoryear{{Myers}, {Shanks}, {Outram}, {Frith} \&
  {Wolfendale}}{{Myers} et~al.}{2004}]{myers_etal04}
{Myers} A.~D.,  {Shanks} T.,  {Outram} P.~J.,  {Frith} W.~J.,    {Wolfendale}
  A.~W.,  2004, MNRAS, 347, L67, \eprint{astro-ph/0306180}

\bibitem[\protect\citeauthoryear{{Nagai}}{{Nagai}}{2006}]{nagai06}
{Nagai} D.,  2006, ApJ, 650, 538, \eprint{astro-ph/0512208}

\bibitem[\protect\citeauthoryear{{Nagai}, {Kravtsov} \& {Vikhlinin}}{{Nagai}
  et~al.}{2007}]{nagai_etal07b}
{Nagai} D.,  {Kravtsov} A.~V.,    {Vikhlinin} A.,  2007, ApJ, in preparation

\bibitem[\protect\citeauthoryear{{Nagai}, {Vikhlinin} \& {Kravtsov}}{{Nagai}
  et~al.}{2007}]{nagai_etal07}
{Nagai} D.,  {Vikhlinin} A.,    {Kravtsov} A.~V.,  2007, ApJ, 655, 98,
  \adsurl{http://adsabs.harvard.edu/cgi-bin/nph-bib_query?bibcode=2007ApJ...65%
5...98N&db_key=AST}, \eprint{astro-ph/0609247}

\bibitem[\protect\citeauthoryear{{Navarro}, {Frenk} \& {White}}{{Navarro}
  et~al.}{1995}]{navarro_etal95}
{Navarro} J.~F.,  {Frenk} C.~S.,    {White} S.~D.~M.,  1995, MNRAS, 275, 720,
  \adsurl{http://adsabs.harvard.edu/cgi-bin/nph-bib_query?bibcode=1995MNRAS.27%
5..720N&db_key=AST}, \eprint{astro-ph/9408069}

\bibitem[\protect\citeauthoryear{{Navarro}, {Frenk} \& {White}}{{Navarro}
  et~al.}{1997}]{nfw}
{Navarro} J.~F.,  {Frenk} C.~S.,    {White} S.~D.~M.,  1997, ApJ, 490, 493,
  \adsurl{http://adsabs.harvard.edu/cgi-bin/nph-bib_query?bibcode=1997ApJ...49%
0..493N&db_key=AST}, \eprint{astro-ph/9611107}

\bibitem[\protect\citeauthoryear{{Ostriker}, {Bode} \& {Babul}}{{Ostriker}
  et~al.}{2005}]{ostriker_etal05}
{Ostriker} J.~P.,  {Bode} P.,    {Babul} A.,  2005, ApJ, 634, 964,
  \eprint{astro-ph/0504334}

\bibitem[\protect\citeauthoryear{{Peres}, {Fabian}, {Edge}, {Allen},
  {Johnstone} \& {White}}{{Peres} et~al.}{1998}]{peres98}
{Peres} C.~B.,  {Fabian} A.~C.,  {Edge} A.~C.,  {Allen} S.~W.,  {Johnstone}
  R.~M.,    {White} D.~A.,  1998, MNRAS, 298, 416, \eprint{astro-ph/9805122}

\bibitem[\protect\citeauthoryear{{Rasia}, {Ettori}, {Moscardini}, {Mazzotta},
  {Borgani}, {Dolag}, {Tormen}, {Cheng} \& {Diaferio}}{{Rasia}
  et~al.}{2006}]{rasia_etal06}
{Rasia} E.,  {Ettori} S.,  {Moscardini} L.,  {Mazzotta} P.,  {Borgani} S.,
  {Dolag} K.,  {Tormen} G.,  {Cheng} L.~M.,    {Diaferio} A.,  2006, MNRAS,
  369, 2013, \eprint{astro-ph/0602434}

\bibitem[\protect\citeauthoryear{{Sanderson}, {Ponman} \&
  {O'Sullivan}}{{Sanderson} et~al.}{2006}]{sanderson06}
{Sanderson} A.~J.~R.,  {Ponman} T.~J.,    {O'Sullivan} E.,  2006, MNRAS, 372,
  1496, \eprint{astro-ph/0608423}

\bibitem[\protect\citeauthoryear{{Schmidt}, {Allen} \& {Fabian}}{{Schmidt}
  et~al.}{2004}]{2004MNRAS.352.1413S}
{Schmidt} R.~W.,  {Allen} S.~W.,    {Fabian} A.~C.,  2004, MNRAS, 352, 1413,
  \adsurl{http://adsabs.harvard.edu/cgi-bin/nph-bib_query?bibcode=2004MNRAS.35%
2.1413S&db_key=AST}, \eprint{astro-ph/0405374}

\bibitem[\protect\citeauthoryear{{Sommer-Larsen}, {Romeo} \&
  {Portinari}}{{Sommer-Larsen} et~al.}{2005}]{larsen_etal05}
{Sommer-Larsen} J.,  {Romeo} A.~D.,    {Portinari} L.,  2005, MNRAS, 357, 478,
  \adsurl{http://adsabs.harvard.edu/cgi-bin/nph-bib_query?bibcode=2005MNRAS.35%
7..478S&db_key=AST}, \eprint{astro-ph/0403282}

\bibitem[\protect\citeauthoryear{{Spergel} et~al.,}{{Spergel}
  et~al.}{2006}]{spergel06}
{Spergel} D.~N.,  et~al., 2006, ArXiv Astrophysics e-prints,
  \adsurl{http://adsabs.harvard.edu/cgi-bin/nph-bib_query?bibcode=2006astro.ph%
..3449S&db_key=PRE}, \eprint{astro-ph/0603449}

\bibitem[\protect\citeauthoryear{{Sunyaev} \& {Zel'dovich}}{{Sunyaev} \&
  {Zel'dovich}}{1972}]{sunyaev72}
{Sunyaev} R.~A.,  {Zel'dovich} Y.~B.,  1972, Comments on Astrophysics and Space
  Physics, 4, 173,
  \adsurl{http://adsabs.harvard.edu/cgi-bin/nph-bib_query?bibcode=1972CoASP...%
4..173S&db_key=AST}

\bibitem[\protect\citeauthoryear{{Vikhlinin} et~al.,}{{Vikhlinin}
  et~al.}{2006}]{vikhlinin_etal06}
{Vikhlinin} A.,  et~al., 2006, ApJ, 640, 691,
  \adsurl{http://adsabs.harvard.edu/cgi-bin/nph-bib_query?bibcode=2006ApJ...64%
0..691V&db_key=AST}, \eprint{astro-ph/0507092}

\bibitem[\protect\citeauthoryear{{Vikhlinin}, {Forman} \& {Jones}}{{Vikhlinin}
  et~al.}{1999}]{1999ApJ...525...47V}
{Vikhlinin} A.,  {Forman} W.,    {Jones} C.,  1999, ApJ, 525, 47,
  \adsurl{http://adsabs.harvard.edu/cgi-bin/nph-bib_query?bibcode=1999ApJ...52%
5...47V&db_key=AST}, \eprint{astro-ph/9905200}

\bibitem[\protect\citeauthoryear{{Voit}}{{Voit}}{2005}]{voit_etal05}
{Voit} G.~M.,  2005, Reviews of Modern Physics, 77, 207,
  \adsurl{http://adsabs.harvard.edu/cgi-bin/nph-bib_query?bibcode=2005RvMP...7%
7..207V&db_key=PHY}, \eprint{astro-ph/0410173}

\bibitem[\protect\citeauthoryear{{White}, {Jones} \& {Forman}}{{White}
  et~al.}{1997}]{white97}
{White} D.~A.,  {Jones} C.,    {Forman} W.,  1997, MNRAS, 292, 419,
  \eprint{astro-ph/9707269}

\bibitem[\protect\citeauthoryear{{Zakamska} \& {Narayan}}{{Zakamska} \&
  {Narayan}}{2003}]{zakamska_etal03}
{Zakamska} N.~L.,  {Narayan} R.,  2003, ApJ, 582, 162,
  \eprint{astro-ph/0207127}

\bibitem[\protect\citeauthoryear{{Zibetti}, {White}, {Schneider} \&
  {Brinkmann}}{{Zibetti} et~al.}{2005}]{zibetti_etal05}
{Zibetti} S.,  {White} S.~D.~M.,  {Schneider} D.~P.,    {Brinkmann} J.,  2005,
  MNRAS, 358, 949,
  \adsurl{http://adsabs.harvard.edu/cgi-bin/nph-bib_query?bibcode=2005MNRAS.35%
8..949Z&db_key=AST}, \eprint{astro-ph/0501194}

\end{thebibliography}

\label{lastpage}
\end{document}